\documentclass[12pt]{article}

\usepackage{amsmath,amsfonts,amssymb,color,graphicx,overpic,} 
\usepackage{slashed}
\usepackage{epsfig}
\usepackage[utf8]{inputenc}
\usepackage{amsmath}
\usepackage{ wasysym }
\usepackage{graphicx}
\usepackage[english]{babel}
\usepackage{graphic x}
\usepackage{relsize}
\usepackage{cite}

\begin{document}

\begin{titlepage}
\rightline{June 2016}
\vskip 2cm
\centerline{\Large \bf
Mirror dark matter 
}
\vskip 0.5cm
\centerline{\Large \bf 
will be confirmed or excluded by XENON1T}

\vskip 2.2cm
\centerline{\bf J. D. Clarke\footnote{E-mail address: j.clarke5@pgrad.unimelb.edu.au} and R. Foot\footnote{
E-mail address: rfoot@unimelb.edu.au}
}

\vskip 0.7cm
\centerline{\it ARC Centre of Excellence for Particle Physics at the Terascale,}
\centerline{\it School of Physics, University of Melbourne,}
\centerline{\it Victoria 3010 Australia}
\vskip 2cm
\noindent
Mirror dark matter, where dark matter resides in a hidden sector exactly isomorphic to the standard model,
can be probed via direct detection experiments 
by both nuclear and electron recoils if the kinetic mixing interaction exists. 
In fact, the kinetic mixing interaction appears to be a prerequisite for consistent small scale structure:
Mirror dark matter halos around spiral galaxies are dissipative - losing energy via dark photon emission.
This ongoing energy loss requires a substantial energy input, which can be sourced from ordinary supernovae
via
kinetic mixing induced processes in the supernova's core.
Astrophysical considerations thereby give a lower
limit on the kinetic mixing strength, 
and indeed lower limits on both nuclear and electron recoil rates in direct detection
experiments can be estimated. 
We show here that 
potentially all of the viable parameter space will be probed in
forthcoming XENON experiments including LUX and XENON1T.
Thus, we anticipate that these experiments will provide a definitive test of the mirror dark matter
hypothesis.

\end{titlepage}


\section{Introduction}

Direct detection experiments have been making remarkable progress in
the search for elusive interactions between dark matter and standard matter.
The discovery potential of 
the large scale xenon experiments probing nuclear recoils, 
including LUX \cite{lux1}, XENON100 \cite{xenon100}, and XENON1T \cite{xenon1T}, 
has rapidly improved in recent years.
In particular, the XENON1T experiment will start a first science run this year,
and is expected to probe dark matter -- nucleon cross sections 
down to the unprecedented level of $\sigma_n \approx 10^{-47} \ {\rm cm^2}$
(for $m_{\rm WIMP} = 50$~GeV)
after two years \cite{xenon1T}.
Meanwhile, 
the DAMA \cite{dama1} and DAMA/LIBRA scintillator crystal experiments \cite{dama2} 
have observed annual modulation in their 
event rate with the characteristic features expected of a dark matter signal. 
The DAMA experiments are sensitive to both nuclear and electron recoils,
but an interpretation involving only nuclear recoils is strongly disfavored
by other experiments (including the xenon experiments listed above).

The physical nature of dark matter is {\it the} critical issue.
Theoretically there are a range of possibilities. 
One well studied and simple possibility is that
dark matter is composed of weakly interacting massive particles (WIMPs). 
Alternatively, more complex scenarios are possible. 
A particularly non-trivial case is plasma dark matter:
a multi-component scenario wherein the dark matter
is minimally composed of dark electrons and dark protons, 
with self-interactions mediated by a massless dark photon,
and taking the form of a dark plasma in spiral galaxies like the Milky Way.
This scenario implies rather unique implications for direct detection experiments.
First, energy equipartition and dark charge neutrality
imply that the standard halo model for single component dark matter
is not at all applicable.
Light component velocities are boosted,
so that e.g. MeV mass-scale dark electrons are energetic enough 
to produce electron recoils in the keV energy range
\cite{footer}.\footnote{Energy equipartition implies that the mean kinetic energy of a light dark electron
component is
identical to that of the much heavier dark proton component. Of course this is only true in the reference
frame stationary with respect to the halo.
In the Earth reference frame (relevant for direct detection experiments) the heavy component still has
greater mean energy
as the galactic rotational velocity, $v_{rot} \approx 220$ km/s,
is much larger [smaller] than the dark proton's [dark electron's] mean thermal velocity.}
Second, long-range self-interactions of the plasma dark matter wind 
with Earth-captured dark matter
leads to spatial variation of the near-Earth dark matter distribution,
macroscopically described by various magnetohydrodynamical features such as bow and tail shocks \cite{jc}.
Annual and sidereal modulation fractions can be strongly enhanced due to these interactions.
Hence plasma dark matter appears to have the potential to explain 
the DAMA annual modulation via electron recoils.

Mirror dark matter (MDM) is the most theoretically constrained 
particle physics model for plasma dark matter \cite{m1,m2,m3,m4,m5,m6,sph}
(for a review and more detailed bibliography see Ref.~\cite{mdmreview}).
MDM postulates a hidden sector which is isomorphic to the ordinary sector.
This means that the dark sector is populated by 
a spectrum of ``mirror'' particles:
$e'$, $H'$, $He'$, $O'$, $Fe'$, etcetera. 
The relevant particle physics is described by just one new (dimensionless) parameter, $\epsilon$, 
which defines the kinetic mixing strength of photons and mirror photons. 
This parameter is restricted from various astrophysical and cosmological
arguments 
\cite{mdmreview,sunny1,sunny2}
to lie within a fairly narrow range, here taken conservatively to be: 
$10^{-11} \lesssim  \epsilon \lesssim 4 \times 10^{-10}$. 

The MDM plasma is expected to be dominantly made up of mirror electrons ($e'$) and mirror helium ($He'$).
However there must be also a non-zero (and constrained)
component of atomic mirror metals.
These may serve as an extra handle on the MDM scenario,
since they are heavy enough to induce detectable nuclear recoils
in the xenon experiments.
Moreover, predictions for their scattering rates are expected to be under better theoretical control, since
existing experiments can probe the body of the velocity distribution
rather than the tail, as occurs for mirror electron -- electron scattering.
In this paper we show that
forthcoming XENON experiments, such as LUX and XENON1T,
will provide a very sensitive probe of MDM 
in both the nuclear and electron recoil channels.
In fact, we argue here that these experiments should be sensitive enough 
to probe essentially all of the parameter
space of interest for MDM.
LUX and XENON1T, together with many other direct detection experiments, are expected therefore to provide a
definitive test of the MDM hypothesis.

\section{Mirror dark matter in the Galaxy}

A generic possibility is that dark matter 
arises from a hidden sector of particles and forces 
coupling to ordinary matter via gravity
and possibly small non-gravitational interactions:
\begin{eqnarray}
{\cal L} = {\cal L}_{SM} + {\cal L}_{dark} + {\cal L}_{mix}
\ .
\label{1x}
\end{eqnarray}
MDM corresponds to the theoretically special case where the hidden sector is isomorphic to 
the standard model sector. 
This ``mirror sector'' has gauge
symmetry $SU(3)'\otimes SU(2)'_L \otimes U(1)'_Y$, 
and the full Lagrangian of Eq.~(\ref{1x})
features an exact and unbroken $\mathbb{Z}_2$ symmetry.
The fundamental properties of the mirror sector 
are therefore completely analogous to that of the standard sector.
The $\mathbb{Z}_2$ symmetry, which can be interpreted as space-time parity \cite{flv},
interchanges each standard particle 
with a corresponding mirror partner, denoted with a prime ($'$).
At the most fundamental level, this implies a duplicate set of elementary particles:
mirror quarks, mirror leptons, mirror gauge bosons, and a mirror Higgs particle.
For astrophysical situations, we are more interested in mirror electrons and the resulting spectrum of 
mirror nuclei:
$e'$, $H'$, $He'$, $O'$, $Fe'$, etcetera,
whose masses  are equal to those of the corresponding standard particle (electron/nuclei).\footnote{%
Although their masses
are fixed their relative abundance are unknown.
Mirror helium synthesis is expected to occur in the early Universe \cite{m3} and can be estimated under some 
simple assumptions \cite{paolo2}, however the heavier ``mirror metal'' components are anticipated
to be synthesised in mirror stars at an early epoch and are much more uncertain (cf.~\cite{m9x}).}

The two gauged $U(1)$ symmetries, $U(1)_Y$ and $U(1)'_Y$,
can kinetically mix \cite{foothe}, 
thereby inducing a small ordinary electric charge for the mirror particles \cite{holdom}.
We parameterise this charge by $\epsilon$
such that e.g. mirror electrons have a charge $- \epsilon e$
and mirror nuclei 
with atomic number $Z'$ have charge $+ \epsilon Z' e$.
It is this interaction which could potentially mediate 
MDM scattering in direct detection experiments.
For a more extensive discussion of the MDM scenario, 
including astrophysical and cosmological implications, 
see the review Ref.~\cite{mdmreview} 
and references therein.

Mirror dark matter is dissipative on time scales relevant for galaxy formation.
The dynamics of dark halos formed in such
dissipative dark matter scenarios 
has been studied in the context of MDM \cite{sph,tf,mdmreview} 
and more generally \cite{sunny1}.
At the current epoch,
the dark matter forms a pressure supported multi-component plasma with both heating and cooling processes.
In particular, the MDM halo consists of an approximately spherical dark plasma 
containing $e'$, $H'$, $He'$, $O'$, $Fe'$, etcetera.
This plasma is expected to evolve to a steady state configuration which is in 
approximate hydrostatic equilibrium and where heating and cooling rates balance.
In this picture, a substantial heat source -- assumed to be ordinary supernovae -- is required
to offset the energy loss due to dissipation.
Kinetic mixing induced processes convert
a fraction of a supernova's core collapse energy into 
$e', \bar e', \gamma'$ \cite{raf,sph,sil}. 
The mirror photons ultimately escape from the region around the supernova 
and are absorbed by the halo
via dark photoionisation of mirror metal components. 
Note that photoionisation can occur since the mirror metal components $m_{A'}\ge m_{O'}$
are not fully ionised in the Milky way,
as the temperature of the Milky Way halo is expected to be
\begin{align}
 T \approx \frac{\bar m v_{rot}^2}{2} \sim 0.3 \text{ keV}, \label{temp}
\end{align}
where $\bar m$ is the mean mass: $\bar m \equiv \sum n_i m_i/\sum n_i$ \cite{sph,mdmreview}.
It is estimated that the total energy absorbed
by the metal component 
in the halo is proportional to $\epsilon^2 \xi_{A'}$ \cite{tf,mdmreview},
where $\xi_{A'}$ is the halo mass fraction of species $A'$.
Here $A'$ can be any metal, i.e. from $A' = O'$ to $A' = Fe'$.
These considerations 
can, in principle, give a lower bound on $\epsilon \sqrt{\xi_{A'}}$ and $\epsilon$. 
This is important for direct detection experiments, 
as the nuclear (electron) recoil rate
is proportional to $\epsilon^2 \xi_{A'}$ ($\epsilon^2$) and thus  
the existence of lower limits on these quantities implies that
rates of these recoils cannot be arbitrarily small.
Below we briefly summarise the astrophysically motivated bounds on $\epsilon \sqrt{\xi_{A'}}$ and $\epsilon$.
Readers interested in more details can consult the literature.

The MDM in the Milky Way 
is presumed to have evolved in time, and is currently
in an approximately steady state configuration where heating and cooling rates balance.
Assuming for simplicity that the metal component is dominated by a single element,
the matching of 
the heating rate to the cooling rate gives an estimate for 
$\epsilon\sqrt{\xi_{A'}} \sim 10^{-10} - 10^{-9}$
(see Eq.~(159) of Ref.~\cite{mdmreview}).
Of course, 
there are significant uncertainties in this estimation, 
and making some allowance for this uncertainty suggests what we believe is a conservative lower limit:
$\epsilon \sqrt{\xi_{A'}} \gtrsim   10^{-11}$.\footnote{%
Uncertainties arise from modelling of supernovae, modelling of the energy spectrum of mirror photons produced
in the region around the supernovae,
modelling of cooling and heating rates, simplifying assumptions for the halo, etc.
}
One could use similar arguments to give a rough estimate for the upper bound on this quantity,
but a more stringent upper bound comes from cosmological arguments \cite{sunny2} 
(for earlier work see also Refs.~\cite{m4,m5}): $\epsilon \lesssim   4\times 10^{-10}$.
Thus, since $\xi_{A'} \le 1$, 
the identified range for the quantity of interest for nuclear MDM scattering can be written:
\begin{eqnarray}
 10^{-11} \lesssim \epsilon \sqrt{\xi_{A'}} \lesssim  4\times 10^{-10} \ . 
 \label{eqepsilonxibounds}
\end{eqnarray}
Given that $\xi_{A'} \le 1$, there is a corresponding lower limit on $\epsilon$:
$\epsilon \gtrsim 10^{-11}$.
So we can also write,
\begin{eqnarray}
 10^{-11}\lesssim \epsilon \lesssim 4\times 10^{-10} \ .
 \label{eqepsilonbounds}
\end{eqnarray}
Again, this is expected to be a conservative lower limit being around an order of magnitude below estimations
from Refs.~\cite{mdmreview,sunny1}.

\section{Direct detection}

The scattering rate of dark matter particles on ordinary matter depends  
first on the local distribution.
In the following we assume a highly simplified model where 
the velocity probability distribution function for each particle species 
in the Earth frame is described by a boosted Maxwellian,
and allowing for
a possible cutoff: 
\begin{eqnarray}
f({\textbf{v}}; {\textbf{v}}_E) \propto e^{-({\textbf{v}}_E + {\textbf{v}})^2/v_0^2} \ \Theta \left( v_{c} -
|{\textbf{v}}_E + {\textbf{v}}|\right),
\end{eqnarray}
where $v_0$ characterizes the velocity dispersion,
${\textbf{v}}_E$ is the
velocity of the Earth with respect to the halo frame,
and the function is appropriately normalised.
As we have already described,
the multi-component MDM halo is in hydrostatic equilibrium
with $T \approx \bar m v_{rot}^2/2$.
For MDM,  $\bar m$, is not a free parameter, 
but constrained to be approximately $\bar m \approx 1.1 $ GeV
from mirror BBN calculations\cite{paolo2}.
The net effect is a mass-dependent velocity dispersion characterized by: 
$v_0 = \sqrt{\bar m/m_{A'}} v_{rot}$. 
Note that, for the mirror metal components, 
$v_0$ is much smaller than the corresponding value for WIMPs, 
taken to be $v_0 = v_{rot}$ in the standard halo model.
In the numerical work, we adopt the standard reference values:
$v_{rot} = 220$ km/s,
and $\langle {\textbf{v}}_E \rangle =  v_{rot} + 12$ km/s 
(which takes into account the Sun's peculiar velocity).

Next we require the cross section for dark matter particles scattering off target nuclei.
In our notation
the kinetic mixing induced electric charge 
of a mirror nucleus, $A'$,
with atomic number $Z'$, 
is $\epsilon Z'e$.
Given it is electrically charged a
mirror nucleus can elastically scatter off an ordinary nucleus, $A$, with
atomic number $Z$ --- essentially Rutherford scattering.
If an incoming mirror nucleus with velocity $v$ scatters off an ordinary nucleus assumed initially at rest,
then the ordinary nucleus would recoil with energy, $E_R$, and
\begin{eqnarray}
\frac{d\sigma_{A'}}{dE_R} = \frac{2\pi \epsilon^2 Z^2 Z'^2 \alpha^2 F^2_A F^2_{A'}}{m_A E_R^2 v^2}\ ,
\label{cs2}
\end{eqnarray}
where $F_A$ ($F_{A'}$) is the form factor 
of the nucleus (mirror nucleus). 
A simple analytic expression for
the form factor, which we adopt in our numerical work, is the one
given by Helm \cite{helm,smith}:
\begin{eqnarray}
F_X  = 3\frac{j_1 (qr_X)}{qr_X} e^{-(qs)^2/2}\ ,
\end{eqnarray}
where $j_1$ is the spherical
Bessel function of index 1, 
$r_X = 1.14 X^{1/3}$ fm is the effective nuclear radius, 
$s = 0.9$ fm,
and $q = (2m_A E_R)^{1/2}$ is the momentum transfer.\footnote{%
Unless otherwise specified,
we use natural units, $\hbar = c = 1$ throughout.}
In Sec.~\ref{SecXeDD} we will find it useful to map the 
MDM metal components onto their ``WIMP equivalent.'' 
Thus we also give the relevant WIMP (here denoted by $\chi$) cross section,
\begin{eqnarray}
\frac{d\sigma_{\chi}}{dE_R} = \frac{m_{A}}{2v^2} \frac{\sigma_n}{\mu^2_{n}}A^2 F^2_{A} \ ,
\label{obama}
\end{eqnarray}
where $\mu_n$ is the $\chi$-neutron reduced mass, 
$\sigma_n$ is the $\chi$-neutron cross-section,
and $A$ is the mass number of the target nuclei 
(the standard assumption of isospin invariant $\chi$-nucleon interactions
has been made).

The scattering rate 
of a dark matter particle off an ordinary nucleus is then:
\begin{eqnarray}
\frac{dR_{\rm DM}}{dE_R} = N_T n_{\rm DM} 
\int_{|{\textbf{v}}| > v_{min}}
\frac{d\sigma_{\rm DM}}{dE_R}
\ f({\textbf{v}}; {\textbf{v}}_E)\ |{\textbf{v}}| \ d^3 v\ ,
\label{55}
\end{eqnarray}
where 
$N_T$ is the number of target nuclei per kg of detector,
$n_{\rm DM}$ is the number density of the scattering dark matter,
and 
$\ v_{min} \ = \ \sqrt{ (m_{A} + m_{\rm DM})^2 E_R/2 m_{A} m^2_{\rm DM} }$.\ \footnote{%
The integration can be performed analytically,
with the result written as a function of $v_0$, 
$\langle {\textbf{v}}_E \rangle$, and $v_{esc}$ 
(see e.g. Ref.~\cite{gondolo}).}
For the mirror metal components, $n_{\rm DM} = \rho_{\rm DM} \xi_{A'}/m_{A'}$,
where $\rho_{dm} \approx 0.3 \  {\rm GeV/cm}^3$
(recall $\xi_{A'}$ is the halo mass fraction
of species $A'$, generally assumed to be unity in single component dark
matter models).
For mirror dark matter the cross section is given in Eq.~(\ref{cs2}) and $v_{c} \to \infty$, 
while for WIMPs Eq.~(\ref{obama}) is
used and we take the cutoff to be the galactic escape velocity $v_{esc} = 544$ km/s.

Mirror dark matter is also an example of a plasma dark matter model which features keV electron recoils
arising 
from mirror electron scattering off atomic electrons. This is kinematically possible due to energy
equipartition between the 
light mirror electron and heavy mirror nuclei halo
components.
Coulomb scattering of a mirror electron off an electron is a spin-independent process  
with cross section:
\begin{eqnarray}
\frac{d\sigma_{e'}}{dE_R} = \frac{\lambda}{E_R^2 v^2} \ ,
\label{cs}
\end{eqnarray}
where
$\lambda \equiv 2\pi \epsilon^2 \alpha^2/m_e$,
and $E_R$ is the recoil energy of the scattered electron, 
approximated as being free and at rest
relative to the incoming mirror electron of speed $v$.
Naturally this approximation can only be valid for the loosely bound atomic electrons, i.e. those with
binding energy much less
than $E_R$.
For a xenon target, it is estimated that there are approximately $44$ such loosely bound atomic electrons per
atom.

In addition to nuclear recoils, 
the xenon experiments such as XENON100 \cite{x1,x2}, XMASS \cite{xmass}, LUX \cite{lux1}, XENON1T
\cite{xenon1T} are
sensitive also to keV electron recoils.
The total event rate above a threshold energy, $E_t$, 
can be determined from Eq.~(\ref{55}) with $v_{c}\to \infty$, $v_0\approx \sqrt{2T/m_{e}}$, 
and neglecting small corrections from $\langle\mathbf{v}_E\rangle$ \cite{jc}:
\begin{align}
 R_e = N_T g_T n_{e'} \lambda \left(\frac{2 m_{e}}{\pi T}\right)^\frac12
 \left(
   \frac{e^{-\frac{E_{t}}{T}}}{E_t}  - \frac{\Gamma\left[0,\frac{E_t}{T} \right]}{T}
 \right), \label{EqTotRate}
\end{align}
where $g_T$ is the number of loosely bound electrons per target atom (for xenon, $g_T \approx 44$),
$n_{e'} \approx 0.17 \ {\rm cm^{-3}}$ is an estimate of the halo mirror electron number density,
and $\Gamma[0,z]$ is the upper incomplete Gamma function.
The temperature, $T$, is the electron temperature, which far from the Earth is expected to be approximately
the
same as the mirror nuclei temperature: $T \sim 0.3$ keV from Eq.~(\ref{temp}) 
(this assumes that all species come to approximate thermal equilibrium).
The interaction of the halo wind with the Earth bound dark matter can significantly modify the halo mirror
electron temperature at the
Earth's location, and the electron interaction rate is especially sensitive to variation in this temperature
(the temperature
is uncertain by around a factor of two or so) \cite{jc}.

\section{Implications for XENON1T \label{SecXeDD}}

The implications of MDM for direct detection experiments have been explored over the
past decade or so \cite{dd1,dd2,dd3,dd4,dd5,dd6}.
The focus of much of this work has been in trying to explain existing anomalies, especially the DAMA annual
modulation signal \cite{dama1,dama2}.
It was realised eventually that MDM, and plasma dark matter in general, is much more
complex than WIMP dark matter. 
New sources of annual modulation are possible arising from the interaction of the halo wind
with the captured DM in the Earth.
The system has recently been explored using magnetohydrodynamic equations \cite{jc}. 
It was found that large annual modulation fraction is typical, especially for electron recoils. 
Additionally, large sidereal day modulation is expected.
Here, we focus primarily on the nuclear recoil spectrum, which is the subject of a number of
large scale experiments, such as LUX and XENON1T. 
Naturally if such a signal were found
in these experiments, annual modulation and sidereal day modulation could also be searched for.
These experiments are also very sensitive to electron recoils, and this will provide a complimentary
way to probe MDM, which will also be briefly considered.

\subsection{Nuclear recoils}

\begin{figure}
 \centering
 \includegraphics[width=0.8\textwidth]{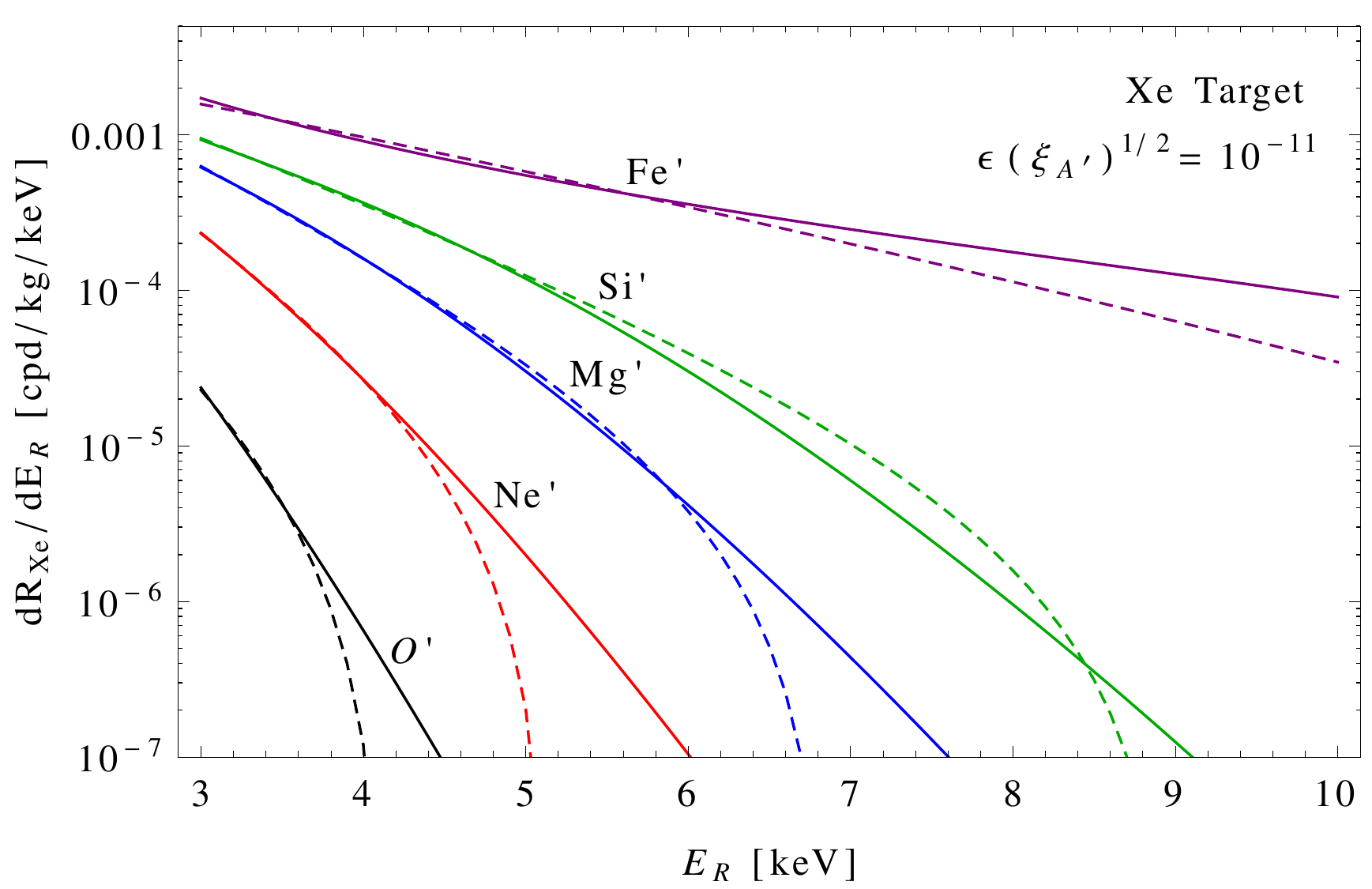}
 \caption{Differential xenon scattering rates
 for a representative set of mirror metals (solid),
 together with their WIMP equivalents (dashed).
 We have taken $v_{rot}^{\rm MDM}=220$~km/s and $\epsilon\sqrt{\xi_{A'}}=10^{-11}$.}
 \label{FigWIMPFits}
\end{figure}

In Fig.~\ref{FigWIMPFits} we show the spectrum, Eq.~(\ref{55}), 
for a representative set of mirror metals scattering on a xenon target. 
Above kinematic thresholds, the rate falls with recoil energy,  
$dR/dE_R \propto 1/E_R^2$, which is due to the $1/E_R^2$ 
dependence of the cross section, Eq.~(\ref{cs2}). 
For a xenon target, the $O'$ ($Fe'$) kinematic threshold is around $E_R \approx  2$ keV ($E_R \approx  10$
keV).\footnote{%
The kinematic threshold for the light mirror helium ($He'$) component ($m_{He'} = 3.7$ GeV) is less than a
keV, which
is below the expected energy threshold  
of these experiments. It might be possible to probe such light dark matter 
in future experiments with light target materials and very low thresholds, for one
such proposal see e.g. \cite{light}.}
The case of lighter target materials, such as germanium,  
is qualitatively similar, with slightly larger threshold energies.

The MDM spectra for a given target element over a sufficiently small energy range 
can be reproduced approximately by a standard WIMP. 
Given that direct detection experiments usually present their results 
as limits on the 
spin independent scattering of
standard halo model WIMPs, we will find it useful to map MDM parameter space
to the ``WIMP equivalent'' parameter space.
To facilitate this we define
the following $L_2$ norm which can be thought of as a continuous least squares fit
between differential MDM and WIMP spectra:
\begin{eqnarray}
L_2 = \int^{E_2}_{E_1} \ \left( \frac{dR_{\rm MDM}}{dE_R} - \frac{dR_{\chi}}{dE_R} \right)^2 \ dE_R
\ .
\end{eqnarray}
Here,
$dR_{\rm MDM}/dE_R$ is the interaction rate for MDM and is a function of 
$A'$, $\epsilon\sqrt{\xi_{A'}}$, and the halo parameters $v_{rot}^{\rm MDM}$, and $\bar m$,
while $dR_{\chi}/dE_R$ is the rate for standard WIMPs with 
reference values fixed to the standard halo model
(we take $v_0 = v_{rot} = 220$ km/s, and  $v_{esc} = 544$ km/s).
For a given energy range, $[E_1, E_2]$,
and MDM parameters, 
we can minimise $L_2$ to obtain the WIMP equivalent $(m_\chi, \sigma_n)$ .
Note that the WIMP equivalent mapping is target-dependent;
it is merely a shape-fitting procedure for the differential rates.
For xenon experiments we choose $[E_1, E_2] = [3 \ {\rm keV_{NR}}, 10 \ {\rm keV_{NR}}]$.
In Fig.~\ref{FigWIMPFits} we illustrate the WIMP equivalents
for a representative set of mirror metals.

%
%
%
%

%
%

\begin{figure}[p]
 \centering
 \includegraphics[width=0.8\textwidth]{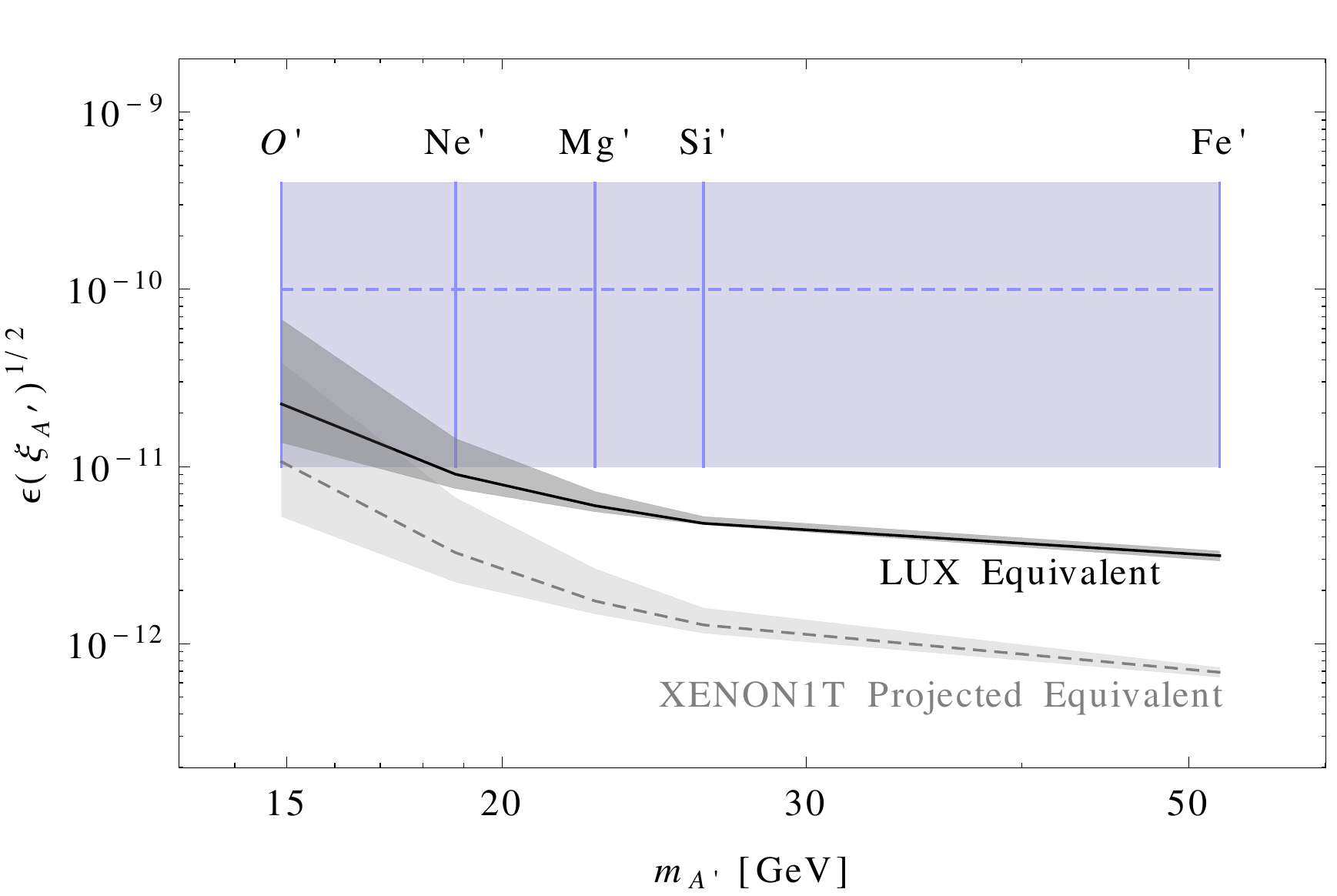}
 \caption{The astrophysically favoured  
 parameter space for a MDM halo with metal component dominated by an element $A'$ from $O'$ to $Fe'$. 
 Also shown are the MDM equivalent LUX and projected 2-year XENON1T exclusions assuming $v_{rot}^{\rm
MDM}=220$~km/s,
 with the band representing the variation for $190\text{ km/s}\lesssim v_{rot}^{\rm MDM}\lesssim 250$~km/s.}
 \label{FigMDMSpace}
\end{figure}

\begin{figure}[p]
 \centering
 \includegraphics[width=0.8\textwidth]{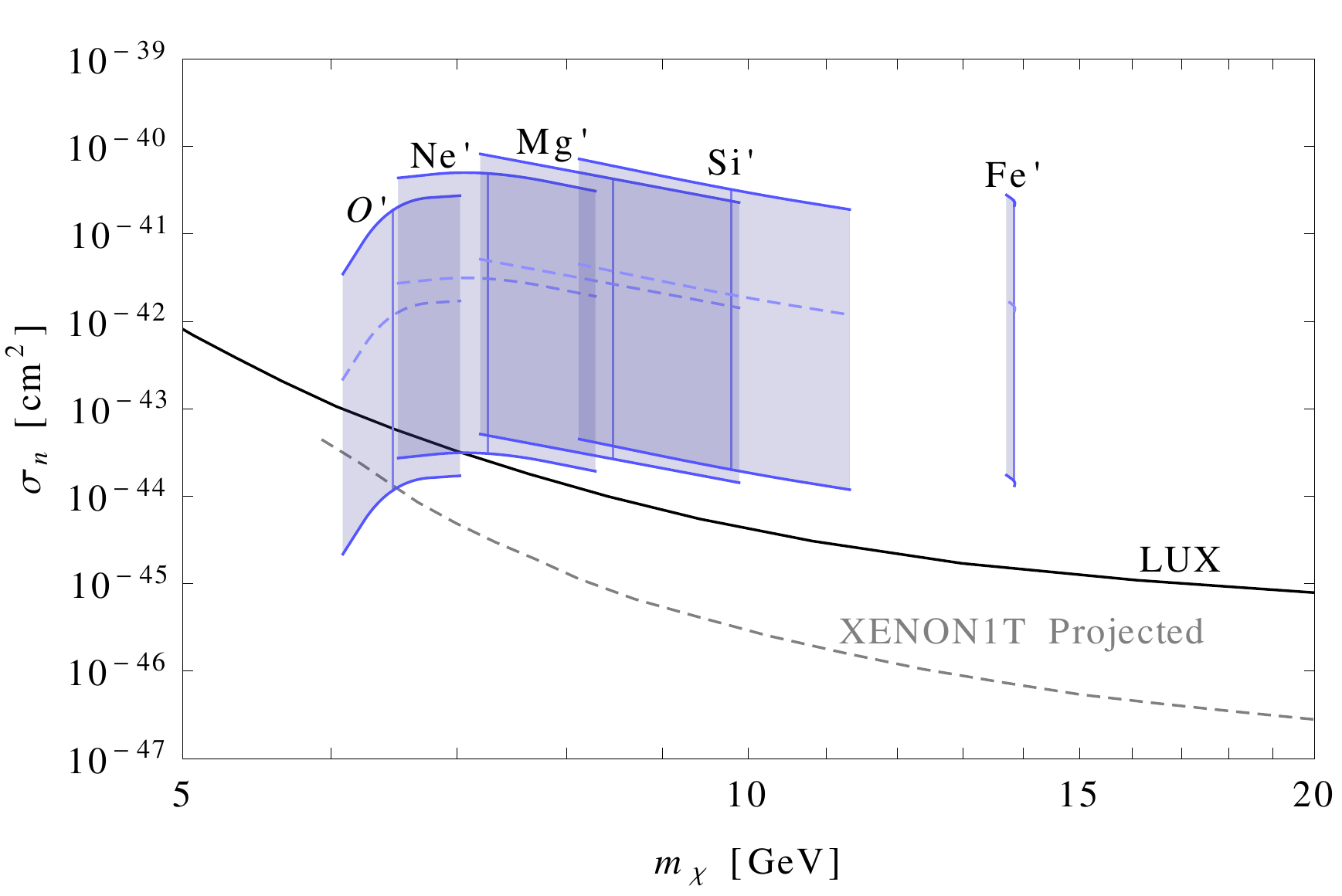}
 \caption{WIMP equivalent parameter space for a MDM halo with dominant metal component as indicated.
 The regions are bounded by $10^{-11}\lesssim \epsilon\sqrt{\xi_{A'}}\lesssim 4\times 10^{-10}$
 (the dashed lines represent $\epsilon\sqrt{\xi_{A'}}=10^{-10}$)
 and $190\text{ km/s}\lesssim v_{rot}^{\rm MDM}\lesssim 250$~km/s,
 with a vertical bar indicating the $v_{rot}^{\rm MDM}=220$~km/s solution.
 The existing LUX exclusion and the projected 2-year XENON1T exclusion are also shown.}
 \label{FigWIMPSpace}
\end{figure}

In the case of a single element, $A'$, dominating the rate, 
the allowed region in the $(m_{A'}, \epsilon \sqrt{\xi_{A'}})$ plane
is shown in Fig.~\ref{FigMDMSpace}.
The LUX 90\% C.L. limit, in terms of the standard WIMP model 
has been given in Ref.~\cite{lux1}.
This limit can be converted to a limit for the MDM model parameters, 
using the mapping procedure outlined above.
In Fig.~\ref{FigMDMSpace} we show the LUX equivalent bound, 
under the assumption of MDM with $v_{rot}^{\rm MDM} = (220\pm 30)$~km/s, 
keeping $\bar m = 1.1$ GeV fixed.
The exclusion curves are not 
marginalised over the energy calibration uncertainty.
This energy scale uncertainty (which is typically around $\sim $ 20\%) is critically important in the low
mass region as
it results in horizontal shift of the exclusion curve, and is thus an important caveat on these bounds.  
Alternatively, we can map the mirror metal regions onto the standard $(m_{\chi}, \sigma_n)$ WIMP parameter
space.
The results of this exercise are shown in Fig.~\ref{FigWIMPSpace}.
We show the WIMP equivalent regions for a selection of mirror metals, 
again keeping $\bar m = 1.1$~GeV fixed while allowing for a range of 
$v_{rot}^{\rm MDM} = (220\pm 30)$~km/s.

Figures~\ref{FigMDMSpace} and \ref{FigWIMPSpace} clearly demonstrate that
having the dominant mirror metal being mirror iron is now excluded by LUX.
The possibility of MDM with mirror oxygen dominating the metal fraction remains viable. 
If the data is analysed in terms of standard WIMPs, then
$m_{\chi} \approx 6$ -- 7~GeV
with $\sigma_n \gtrsim  10^{-44} \ {\rm cm^2}$ for a Xe target is implicated -- a very specific parameter
region.
The expected sensitivity of XENON1T after two years of running \cite{xenon1T} is also shown in
Figs.~\ref{FigMDMSpace} and \ref{FigWIMPSpace}.
This experiment will be able to extensively probe this scenario,
and improve the limits on the mass fraction of heavier mirror metals substantially.
The large scale xenon experiments will be particularly sensitive to the
$Ne', Mg', Si', Fe'$ components as there is less kinematic suppression (cf. $O'$) due to their larger masses. 
This means that these heavier metal components can potentially produce a larger signal in the xenon
experiments even
if they are subdominant.
Although we have focused our discussion on the large scale xenon experiments, many other experiments with
lighter target elements
(e.g. CRESST, superCDMS, Edelweiss, ...) can probe MDM, and may provide a more sensitive probe of the mirror
oxygen component.

\subsection{Electron recoils}

\begin{figure}
 \centering
 \includegraphics[width=0.8\textwidth]{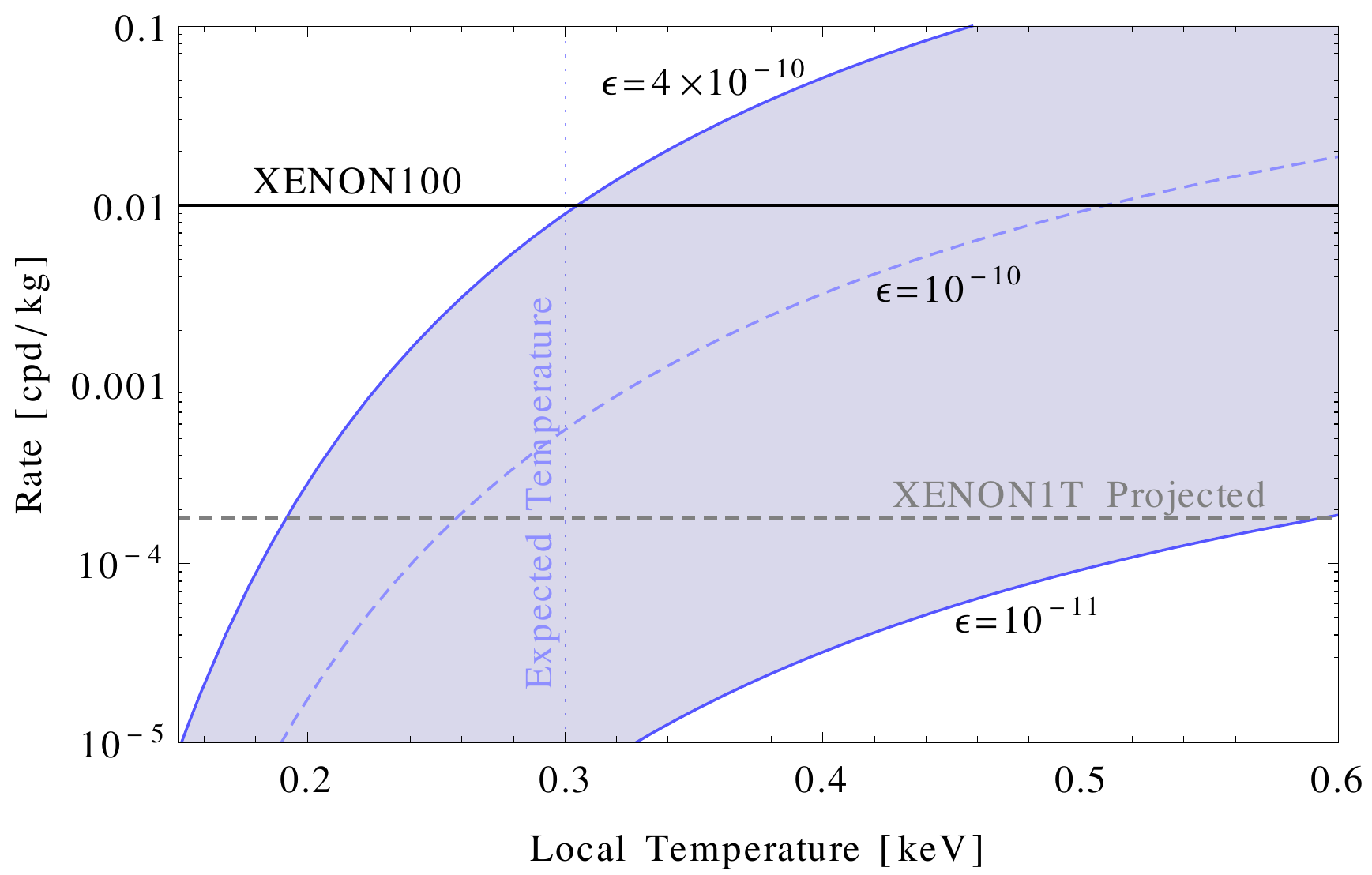}
 \caption{Total rate of electron recoils, Eq.~(\ref{EqTotRate}),
  as a function of the mirror electron temperature at the detector's location 
  assuming a Maxwellian distribution.
  Shown is the rate for a
  xenon detector with threshold energy $E_t = 2$ keV and kinetic mixing
  $10^{-11}\lesssim \epsilon \lesssim 4\times 10^{-10}$. }
  \label{FigElectronRate}
\end{figure}

In Fig.~\ref{FigElectronRate} we show the total rate, Eq.~(\ref{EqTotRate}), 
for a xenon detector as a function of the mirror electron temperature at the detector's location. 
A low energy threshold of $E_t = 2$ keV is assumed.
Also shown in the figure is an estimate of the existing limit 
obtained from Ref.~\cite{x1},
and the estimated reach of XENON1T \cite{xenon1T}.
Fig.~\ref{FigElectronRate} indicates that forthcoming experiments  
will be able to probe 
much of 
the electron recoil parameter space of interest.
We note that the existing limit is already extremely interesting as
it is in some tension with interpretations of the DAMA annual modulation signal in terms of electron recoils.
However, there remain significant uncertainties when different experiments are compared:
at low recoil energies the event rate is the convolution
of a steeply rising event rate,
a generally non-Gaussian resolution,
and a falling efficiency.
DAMA's large resolution can effectively boost 
the electron recoil rate as actual recoils below their $2$ keV 
threshold can in principle dominate their measured rate. 

%

\section{Conclusion}

Hidden sector dark matter allows for rich and diverse possibilities for the dark matter inferred to exist in
our Universe.
A special case arises when the hidden sector is exactly isomorphic to the standard model, so-called mirror
dark matter, which is particularly
theoretically constrained as it contains only one free fundamental parameter, 
the kinetic mixing strength with standard matter, $\epsilon$.
It has been
argued in previous work that kinetic mixing is required to make the resulting dissipative dark matter
galaxy halos consistent with rotation curve measurements.
In this picture, kinetic mixing provides a critical role by transforming ordinary supernovae 
into powerful galactic halo
energy sources which can replace the halo energy lost due to dissipation.
The halo must contain a mirror metal component so that the energy sourced from ordinary supernovae
can be absorbed efficiently 
in the halo, although such astrophysical considerations have not as yet provided any clues as to which
particular mirror metal, $A'$, 
dominates the spectrum (i.e. the possibilities range from $A' = O'$ to $A' = Fe'$). 
Such astrophysical considerations thereby give a lower
limit on the kinetic mixing strength, and indeed lower limits on both nuclear and electron recoil rates in
direct detection
experiments can be estimated.

Examining first nuclear recoils,
we have shown here that 
having the dominant mirror metal being mirror iron is excluded by the LUX experiment,
however the possibility of MDM with mirror oxygen dominating the metal fraction remains viable. 
In fact, the allowed parameter space is constrained by both astrophysical and experimental considerations
to be $A' \sim O'$ with $10^{-11} \lesssim \epsilon \sqrt{\xi_{A'}} \lesssim 10^{-10}$ (Fig. 2).
This parameter space can be transformed into the standard ``WIMP equivalent'' 
of $m_{\chi} \approx 6$ -- 7~GeV
with $\sigma_n \gtrsim  10^{-44} \ {\rm cm^2}$ for a xenon target (Fig. 3).
This rather specific region is expected to be probed in forthcoming XENON experiments including
LUX and XENON1T (and of course also in many other experiments).
In addition, electron recoils were also considered. We found that this channel, which provides a distinctive
signal for MDM (and plasma dark matter in general), may also 
yield a signal in forthcoming experiments such as XENON1T (Fig. 4).
Furthermore any putative MDM signal can be checked for the (expected large) annual and sidereal day
modulations.
Thus, we anticipate that mirror dark matter will be definitively tested by these experiments in the near
future.

\vskip 1cm
\noindent
{\large \bf Acknowledgments}

\vskip 0.1cm
\noindent
This work was supported by the Australian Research Council.
We would also like to thank R. Lang for useful correspondence regarding the xenon direct detection
experiments.

\end{document}